# Electronic structure studies of Fe doped ZnO nanorods by x-ray absorption fine structure

Sanjeev Gautam[1,*], S. Kumar[2], P. Thakur[3], K. H. Chae[1], Ravi Kumar[4], B.H. Koo[2] and C. G. Lee[2]

[1]Nano Materials Analysis Center, Korea Institute of Science and Technology, Seoul 136-791, Republic of Korea

[2]School of Nano & Advanced Materials Engineering, Changwon National University, 9 Sarim dong, Changwon 641-773, Republic of Korea

[3]European Synchrotron Radiation Facility, BP220, 38043, Grenoble Cedex, France

[4] Materials Science Division, Inter University Accelerator Centre, New Delhi 110067, India

Phone: + 82 54 279 1192, Fax: + 82 54 279 1599, E-mail: khchae@kist.re.kr (K.H. Chae); sgautam71@kist.re.kr (S. Gautam)

**Abstract.**

We report the electronic structure studies of well characterized polycrystalline $Zn_{1-x}Fe_xO$ (x = 0.0, 0.01, 0.03, and 0.05) nanorods synthesized by a co-precipitation method through x-ray absorption fine structure (XAFS). X-ray diffraction (XRD) reveals that Fe doped ZnO crystallizes in a single phase wurtzite structure without any secondary phase. From the XRD pattern, it is observed that peak positions shift towards lower 2θ value with Fe doping. The change in the peak positions with increase in Fe contents clearly indicates that Fe ions are replacing Zn ions in the ZnO matrix. Linear combination fittings (LCF) at Fe *K*-edge demonstrate that Fe is in mixed valent state ($Fe^{3+}/Fe^{2+}$) with a ratio of ~7:3 ($Fe^{3+}:Fe^{2+}$). XAFS data is successfully fitted to wurtzite structure using IFEFFIT and Artemis. The results indicate that Fe substitutes Zn site in the ZnO matrix in tetrahedral symmetry.

*PACS*: 75.50.Pp; 18.70.Dm; 61.46. +w; 82.80.Ej

*Keywords*: DMS; ZnO; X-ray diffraction and absorption;

---

[*]Also at Pohang Accelerator Laboratory, Pohang University of Science and Technology, Pohang -790 784, Korea

## 1. Introduction

In the recent years spintronics materials continue to be of the great interest to the scientific community due to their potential technological applications. The basic idea is to combine the characteristics of existing magnetic devices with semiconductor devices in order to realize the new generation of devices that are smaller, energy efficient, and faster than presently available devices [1-9]. The key requirement in the development of such devices is an efficient injection, transfer, and detection of spin-polarized currents at/or above the room temperature (RT). However, due to well known problem of resistance/lattice mismatch at metal/semiconductor interfaces, hindering an effective spin injection [10], much interest is now concentrating on the development of ferromagnetic (FM) semiconductors at RT.

The dilute magnetic semiconductors (DMSs) can be designed by replacing the fraction of the non-magnetic semiconductor cation by the $3d$ transition metal (TM) cations. DMSs are promising candidates for spintronics applications at ambient temperatures, provided that their Curie temperature ($T_C$) is far above the RT. Therefore, a number of different semiconductor hosts have been investigated to test their magnetic properties. In the past, most attention has been paid to (Ga, Mn)As [11-17] and (In,Mn)As [18-23] DMS systems. However, their reported highest $T_C$ are only around 170 K for (Ga,Mn)As [24-25] and 35 K for (In,Mn)As [25-26]. Therefore, there is a large incentive to develop new DMSs with much higher Curie temperature. In particular, the calculations of Dietl *et al.* [27] were first to indicate that Mn doped ZnO could exhibit FM above RT. Later Sato and Katayama-Yoshida also investigated ZnO-based DMS by *ab initio* electronic structure calculations and reported ferromagnetic ordering in $3d$ transition metal (TM) doped ZnO [28-29]. These theoretical predictions initiated an outburst of experimental activities on TM-doped ZnO [30-36]. However some of these studies indeed claim ferromagnetism above RT. But, the origin of ferromagnetic behavior in these systems is still under debate. The main unresolved question is weather the observed ferromagnetism originates from uniformly distributed TM cations in the ZnO host matrix or weather it is due to the precipitation of metallic ferromagnetic clusters or secondary phases.

In the present work, we have studied the effect of Fe doping on the local electronic structure properties of the ZnO by using x-ray absorption fine structure (XAFS) and extended x-ray absorption fine structure (EXAFS) techniques at Fe $K$- and Zn $K$-edges. It is well known that the magnetic properties of the DMSs are very sensitive to the valance state of the TM cations and local 3$d$-3$d$ interactions. Various groups have reported on the magnetic properties of Fe-ZnO, but still there is a controversy about valence state of Fe as well as the origin of ferromagnetism. Wu *et al.* [37] reported that Fe in ZnO matrix exists in a mixed valence state (i.e. $Fe^{2+}$ and $Fe^{3+}$) in case of films and nanoparticles. However Shinagawa *et al.* [38] reported that there is possibility of the formation of nanosized $ZnFe_2O_4$ spinel phase. Recently, our group has reported ferromagnetism in Fe-ZnO system at RT prepared by ion implantation and followed by a swift heavy ion irradiation technique [39]. These results were explained in the light of mixed valent state of Fe ions ($Fe^{2+}/Fe^{3+}$) and hole mediated ferromagnetism as theoretically predicted by Dietl *et al.* [27] and Sato *et al.* [29].

2. Experimental details

Compounds with nominal composition of $Zn_{1-x}Fe_xO$ (x = 0.0, 0.01, 0.03 and 0.05) were synthesized by co-precipitation method. Analytical grade metal nitrates [$Zn(NO_3)_2.6H_2O$, $Fe(NO_3)_3.9H_2O$] were used to prepare Fe-ZnO samples. The details of the synthesis of Fe-ZnO and other magnetic characterizations are reported elsewhere [40]. The samples are in the form of nanorods with perfect crystalline structure and exhibit ferromagnetism at RT [40]. The x-ray diffraction (XRD) measurements were carried out at RT using a Phillips X'pert (MPD 3040) x-ray diffractometer with a Cu $K_\alpha$ source (λ=1.5406Å) operated at a voltage 36 kV and current 30 mA. For XAFS experiments at Fe $K$- and Zn $K$-edges, 10B XRS KIST-PAL beamline of the Pohang Accelerator Laboratory (PAL), operating at 2.5 GeV with a maximum storage current of 200 mA was utilized. This beamline is monochromatized by a double-crystal Si (111) monochromator detuned from 30 to 40%, to suppress higher-order harmonic content from the beam. All the scans were made in transmission mode with nitrogen-argon gas-filled ionization chambers as detectors. The Fe $K$- and Zn $K$-edge energy was calibrated using the first inflection point of the edge region of a metallic Fe/Zn foil. The resolution of the monochromator was about 1.5 eV in the energy range studied. Several scans are taken and averaged to produce a high quality spectrum for analysis. Prior to analysis, the EXAFS spectra were subjected to subtraction of the atomic absorption using the AUTOBK program [41] and corrected for monochromator induced energy shifts using spectra of Fe and Zn foil. The model for fitting was constructed using Athena 0.8.056 [42]. Self absorption of the samples through various algorithms available with Athena was checked and found negligible for all applicable algorithms. All data were collected to 16.0 Å$^{-1}$ at Fe and Zn $K$-edge, but used upto 13.0 Å$^{-1}$ in our calculations to minimize the errors. Fourier transforms (FT) formation and background removal procedures for isolating the oscillatory part of the absorption coefficient were done using the Athena.

## 3. Results and discussion

### 3.1 XRD data analysis

XRD pattern obtained from $Zn_{1-x}Fe_xO$ (x = 0.0, 0.01, 0.03 and 0.05) is shown in Fig. 1. XRD data was analyzed using Powder-X software [43] which reflects that all samples exhibit single phase nature with wurtzite lattice and exclude the presence of secondary phase. In order to see the insight of the effect of Fe doping on the diffraction of doped ZnO, we have done a careful analysis of peaks position in the XRD pattern using (002) plane. From the analysis of the peaks position, it is observed that the peaks position shifted towards lower 2θ value with increasing the Fe contents. This shift in the peak position clearly reflects that Fe is replacing Zn in the ZnO matrix.

**3.2 Near edge x-ray absorption fine structure (NEXAFS) analysis**

In order to study the local electronic structure around Fe and Zn atoms, we have carried out the XAFS/EXAFS experiments at Fe $K$- and Zn $K$-edges. The XAFS/EXAFS techniques have been established as a powerful method to understand the local structure of 3$d$ TM-oxide systems. Photons at characteristic energies are absorbed to produce the transition of a core electron to an empty state above the Fermi level and are governed by dipole selection rules. XAFS is a fingerprint of the electronic state (valence state) of TM cations and is highly sensitive to inspect the presence of TM clusters and other impurities in the host matrix, whereas the quantitative structural information about the metal environment (3$d$-3$d$ coordination state) can be obtained from the EXAFS analysis. Fig. 2(a) show the Fe $K$-edge XAFS spectra of $Zn_{1-x}Fe_xO$ (x = 0.01, 0.03 and 0.05). As a guide of the valence state of Fe ions, reference Fe $K$-edge XAFS spectra of Fe metal foil ($Fe^0$), FeO ($Fe^{2+}$), $Fe_2O_3$ ($Fe^{3+}$), and $Fe_3O_4$ ($Fe^{2+}/Fe^{3+}$) are also shown. It is well known that the peak positions and the spectral line shape of the 3$d$ metal $K$-edge XAFS spectrum depend on the local electronic structure of the metal ions, therefore, provides the information on the valence state of functioning ions in a system. The observed XAFS spectral features of Fe-ZnO are very similar to those of $Fe_3O_4$ (with a positive shift of white line), indicating a mixed-valent state ($Fe^{2+}/Fe^{3+}$) of Fe ions. Previously, we have shown that O K-edge spectra (O1s-2p hybridization) of Fe doped ZnO nanorods [40] do not show any resemblance with any reference compound spectra, therefore, the possibility of $Fe_3O_4/ZnFe_2O_4$ (or $FeO/Fe_2O_3$) precipitation in ZnO matrix below the detectable size limit of XRD is ruled out. It is also noticed that the pre-edge spectral features of the Fe-ZnO (Fe doped ZnO system) are different from the Fe metal foil ($Fe^0$) which exclude the formation of metallic phase in the system. The pre-edge peak near 7114 eV (marked by an arrow) results from $1s$ to $3d$ transition and has a significant intensity because of $3d$-$2p$ orbital mixing. In general, pre-edge centroid position depends strongly on the Fe oxidation state; however the pre-edge intensity is mostly influenced by the Fe coordination geometry. The low intensity of the pre-edge peak refer to the geometries with center of symmetry (e.g., octahedral); whereas the high intensity refer to non-centrosymmetric geometries (e.g., tetrahedral). It can be seen in Fig 2(a) that the intensity of the pre-edge peak has a considerable spectral weight in Fe doped ZnO system as compared to reference compounds and shifted towards higher photon energy side. These changes in the spectral feature of Fe doped ZnO shows that the local electronic structure of Fe ions is different from that of reference compounds of Fe metal foil ($Fe^0$), FeO($Fe^{2+}$), and $Fe_2O_3$($Fe^{3+}$), but bears a close resemblance to $Fe_3O_4$($Fe^{2+}/Fe^{3+}$) spectra. In $Fe_3O_4$, $Fe^{2+}$-$Fe^{3+}$ are equally distributed in octahedral and tetrahedral sites, whereas in case of Fe-ZnO the most of $Fe^{2+}$-$Fe^{3+}$ ions are at tetrahedral site of ZnO matrix (explained below in EXAFS analysis). A shift in photon energy (~1.5 eV) towards higher energy side indicates more $Fe^{3+}$ components in Fe doped ZnO as compared to $Fe_3O_4$, i.e., the presence of more charge carriers (electron/hole), which are responsible for the structural evolution of the local environment of Fe atoms in Fe-ZnO. In order to obtain better inside of the valence state of Fe ions, a linear combination fitting (LCF) of the reference spectra was carried out from -10 eV to 50 eV from the edge energy. Figure 2(b) shows the observed and simulated XAFS spectra for Fe doped ZnO system. It is observed that all the XAFS spectra can be best simulated only when $Fe_3O_4$ ($Fe^{2+}/Fe^{3+}$) spectrum is included, and Fe metal foil ($Fe^0$) is excluded. Exclusion of Fe-foil in the LCF fitting also indicates the absence of metallic phase and/or clusters in the Fe doped ZnO system. In this analysis, both the relative weights and absorptions edge energies are allowed to vary. The energy shifts are within ± 0.5 eV. Table 1 shows the LCF analysis results, which proves the presence of mixed-valent ($Fe^{3+}/Fe^{2+}$) state of Fe ions with a maximum ratio of ~7:3 ($Fe^{3+}$: $Fe^{2+}$). The R-factor and chi-square values are related to the accuracy of the best fit and showing a very small variation of the percentage error in the calculated XAFS spectra. Zn K-edge XAFS spectra does not show any spectral variation for all Fe concentration in ZnO (not shown here), which indicates that the Zn related defects are negligible.

## 3.3 Extended x-ray absorption fine structure (EXAFS) analysis

It is well known that the local structure around a probe atom can be described by studying the fine structure above the absorption edge [44]. In order to know the local structure around the Fe and Zn atom, we have done the EXAFS analysis. The EXAFS spectrum can be understood in the terms of the EXAFS equation. The EXAFS equation can be written in terms of the contribution from all scattering paths of the photoelectron [45-46]:

$$\chi(k) = \sum_i \chi_i(k) \tag{1}$$

Where

$$\chi_i(k) = \frac{(N_i S_0^2) F_{eff_i}(k)}{k R_i^2} \sin[2kR_i + \varphi_i(k)] e^{-2\sigma_i^2 k^2} e^{\frac{-2R_i}{\lambda(k)}} \tag{2}$$

with $R = R_{0i} + \Delta R$ and $k^2 = \frac{2m_e(E - E_0 + \Delta E_0)}{\hbar}$ (3)

Equation (2) is MS (multiple scattering) expansion as a sum over MS paths R, where $k$ = wave number measured from threshold $E_0$, $N_i$= number (coordination number) of atoms of type $i$ at a distance $R_i$ from the absorbing atom, $S_0$ = amplitude factor, $F_{eff}(k)$ = effective scattering amplitude, $\varphi_i$ = phase of the back-scattering factor, $\lambda(k)$= elastic mean free path of the photoelectron and finally $\sigma$, which characterizes the thermal and structural disorder, is the rms (root mean square) fluctuations in the *effective path length* R=R$_{path}$/2, which corresponds to peak in EXAFS Fourier transform, known as Debye-Waller (DW) factor ($\exp(-2\sigma^2 k^2)$). $\sigma_i^2$ is the mean square variation in inter-atomic distances. The absorption coefficient ($\mu^\dagger$) was analyzed and processed by the Athena V0.8.058. EXAFS data were than subjected to theoretical calculations through Artemis V0.8.012 [42] with the IFEFFIT package version 1.2.11c [47].

Figures 3(a) and (b) show the Fe and Zn $k^3$-weighted EXAFS oscillation spectra of Zn$_{1-x}$Fe$_x$O (x = 0.0, 0.01, 0.03, 0.05) along with FeO, Fe$_2$O$_3$, Fe metal foil, and Zn metal foil for comparison. The $k^3\chi(k)$ for the pure and Fe doped ZnO as a function of the photoelectron wave vector $k$ provides the detailed information about the average immediate atomic environment of the given lattice site [48-49]. To minimize the errors, we have considered the data from k range of 2-13 Å$^{-1}$ for further analysis. It is also confirmed that for the Fe concentrations from x = 0.01 to 0.05, the samples have similar $k^3\chi(k)$ curves matching with Fe$_3$O$_4$ curve, which is in agreement with the results of the near edge XAFS. In Fig. 3(a) qualitative similarities with Fe$_3$O$_4$ spectra and effect of Fe doping into the ZnO matrix are clearly visible and are highlighted by vertical dashed lines. In Fig. 3(b) there is no visible effect of addition of Fe into ZnO matrix is seen, implying that Fe is doped at Zn site without the formation of any impurity phase. The Fig. 3 (a) and (b) also proves the good quality single phase sample and XAFS data collected.

---

$^\dagger$ $\mu = \dfrac{1}{x} \ln\left[\dfrac{I_0}{I_t}\right]$, where x=thickness of the sample, I$_t$= transmitted x-ray and I$_0$=x-ray impinging on the sample.

Figures 4(a) and 4(b) show the results of Fourier transform (FT) of EXAFS oscillation $k^3\chi(k)$, representing the radial distribution function (RDF) at Zn $K$- and Fe $K$-edge, respectively. The graphs show the position of atoms (Zn and Fe) which contribute to the scattering wave construction. Several peaks can be clearly observed at a longer radial distance, indicating that the local structure around the Fe atoms are well ordered with respect to short range ordering. In Fig. 4(a), the peak positions located at 1.56 Å and 2.90 Å corresponds to the Zn-O and Zn-Zn bonding distances whereas the main peaks in Fig. 4(b) corresponds to Fe-O (1.93 Å) and Fe-Fe (2.94 Å), respectively. These comparable inter-atomic distances of Fe-Fe and Zn-Zn bonding clearly indicate that Fe atom replaces Zn atom with a small variation of ~0.04 Å. This small variation of ~0.04 Å is also visible at the first peak of Fe-O bonding (marked by an arrow in Fig. 4(b)), indicating that local structure of Fe atoms change with increasing Fe concentration (x = 0.01-0.05). The observed change in the local environment of Fe and Zn atoms is consistent with the EXAFS analysis done by Lin *et al*.[50]. Moreover, the FT curve for $Fe_3O_4$ shows a cubical inverse spinel structure, which is missing in Fe-ZnO. FT curves as shown by arrows pointing $T_d$ (tetrahedral) and $O_d$ (octahedral) symmetries [51] in Fig. 4(b). Hence FT plots show that Fe doped ZnO holds tetrahedral symmetry and the absence of $Fe_3O_4$ related spinel phase such as $ZnFe_2O_4$ . Further, Fe doping seems to affect the two shoulder peaks (as shown in the inset of Fig. 4(a) for x = 0.03, and 0.05 with undoped ZnO) at 0.83 Å and 1.17 Å only, which have become less broad after Fe doping. Shoulder peaks were significant for undoped ZnO (Fig. 5(b)) than that of Fe-ZnO. These shoulder peaks are also discussed by Mu *et al*. [52] and explained on the basis of size of the dopant and steric effects.

To further understand the better insite of the doping effects, the EXAFS data is fitted to wurtzite structure of ZnO using the IFEFFIT code [47]. Background subtraction and data merging was done with Athena [42]. Theoretical models were constructed using the IFEFFIT [47,55] codes and the crystallographic atomic positions of ZnO. After background removal the spectra were Fourier-transformed over a photoelectron wavenumber ($k$) range of 2.0-13.0 Å$^{-1}$. The amplitude reduction factor ($S_0^2$) and energy shift parameter ($\Delta E_0$) are calculated from Zn-foil as: 0.68 ± 0.10 and 5.47±0.60 eV, respectively and kept constant for further calculations. The mean inter-atomic distance (R), variance (Debye-Waller (DW, $\sigma^2$) factor), are determined by individual fits. The $S_0^2$ (amplitude factor) value is low as expected and it may be due to the dead time correction and fluorescence data mode. But as the $S_0^2$ value remains almost constant with addition of paths and R-range, so can be used further for the Fe doped ZnO. Single-scattering (SS) and multiple-scattering (MS) paths were taken into account in the data analysis. It is also noted that below R$_{eff}$ < 4.56 Å only SS paths are used for doped samples, while MS is missing in case of undoped ZnO.

Figure 5 shows the EXAFS data (dotted lines) of pure and Fe-doped ZnO and simulated data (solid lines). The primary purpose of this fit is to demonstrate the efficacy of the fitting model, and therefore, the final results are compared to the nominal crystal structure to quantify any systematic error. To this end, all measured distances are close to those measured by diffraction, although outside the systematic errors. Considering that only 4 fit parameters describe all the bond lengths upto 5 Å, the systematic errors in the pair distances are expected to be within about 0.02 Å, [56] as observed. All the $\sigma^2$ parameters are small, as expected for a well-ordered crystal lattice. The fits to the Zn $K$-edge data from all the samples give similar results, so systematic errors are expected to better than 0.1. However, this error may be smaller when a higher fraction of a particular substituent species resides on the Zn (1) site, as determined.

Table 2 show the fit results for Zn $K$-edge data on pure ZnO and Fe-doped ZnO, which presents the bond length (R), relative Debye-Waller factor ($\sigma^2$), and $S_0^2$ value for X-O and X-X co-ordinations, if X = Fe and Zn. In the EXAFS simulation model, one of the Zn-atom is replaced by Fe substituted ZnO system and taking Fe as the central absorber in the feff.inp file. All scattering paths are included within the fitting range, but only those single-scattering paths with independent pair distances are reported here. All other path distances are constrained to these paths. The data is fitted in the range between 1.0 -6.0 Å in real space and the $k^3$-weighted data is transformed between 2.0-13.0 Å$^{-1}$ after Gaussian narrowing of 0.2 Å. Other parameters and error percentage is shown in Table 3. These fits have about 20 degrees of freedom [57]. Moreover, the data also show a reduction in the amplitude of the peak near r ~ 2.9 Å of 70% compared to the experimental peak, indicating the percentage of Fe atoms residing on Zn site. Parameters like $\sigma^2$ and quality factor (R) are taken care and are good for this fitting.

The bond length and other parameters derived from EXAFS calculations as shown in Table 3 are in consistent with our experimental finding and other published results [52-54]. The bonding length of ZnO obtained from EXAFS experiments fits well to the values calculated by Artemis (see table 2). In case of pure ZnO, the Zn-Zn bonding length is 3.20 Å, which is consistent with the XRD (3.251 Å) results [40,52]. However, the bond lengths of Zn-O and Zn-Zn were found to increase (see Table 2) with Fe doping which are in good agreement of our previous report [40]. The increase of Zn-Zn bonding length might be due to the steric effect. In a wurtzite structured ZnO unit cell, one oxygen atom is neighboring with four Zn atoms forming the geometry of triangular pyramid with oxygen located in the pyramid center. The substitute of Fe atom in place of Zn atom takes less space than a Zn atom and hence Zn-O and Zn-Zn distance should be a bit larger.

**4. Conclusions**

In summary, the electronic structure has been studied using XAFS of $Zn_{1-x}Fe_xO$ (x = 0.0, 0.01, 0.03, and 0.05) polycrystalline nanorods synthesized by a co-precipitation method. XRD measurements reflect single phase polycrystalline nature with wurtzite lattice and exclude the presence of any secondary phase. LCF fitting on XAFS spectra indicates that Fe exists in $Fe^{+3}/Fe^{2+}$ mixed valent state and in tetrahedral symmetry. We have also performed IFEFFT fitting using wurtzite structure of ZnO and calculated Zn-O and Zn-Zn inter-atomic distances and other parameters, which are consistent with values available in literatures. The EXAFS analysis of Fe and Zn *K*-edge reveals that all the Fe ions occupy Zn position in ZnO matrix. It is expected that this study can provide new structural information useful for material design and property optimization of TM doped ZnO DMS materials.

**Acknowledgements**
This work was supported by Korea Institute of Science and Technology (KIST, Grant No. 2V01450) and KRF grant (MOEHRD) KRF-2006-005-J02703).

# Figure Caption

Figure 1: (color online) X-ray diffraction pattern of $Zn_{1-x}Fe_xO$ (x = 0.0, 0.01, 0.03, and 0.05).

Figure 2: (color online) Normalized Fe $K$-edge XAFS Spectra of $Zn_{1-x}Fe_xO$ (x = 0.01, 0.03, and 0.05) at RT: (a) Observed XAFS spectra compared with those of reference compounds of $FeO(Fe^0)$, $Fe_2O_3(Fe^{2+})$, and $Fe_3O_4(Fe^{2+}/Fe^{3+})$; (b) Calculated XAFS spectra of $Zn_{1-x}Fe_xO$ simulated (solid line) by using linear combination fitting of reference compounds plotted with the experimental data (dots).

Figure 3: (color online) Normalized $k^3\chi(k)$ weighted EXAFS spectra of $Zn_{1-x}Fe_xO$ (x = 0.01, 0.03, and 0.05) at RT: (a) Observed XAFS spectra at Fe $K$-edge compared with Fe-foil, FeO, $Fe_2O_3$ and $Fe_3O_4$ reference spectra. (b) Observed XAFS spectra at Zn-$K$ edge for $Zn_{1-x}Fe_xO$ (x = 0.0, 0.01, 0.03, and 0.05) plotted with Zn-foil for comparison. Vertical dashed lines show the spectral differences and similarities in (a) and (b) respectively.

Figure 4: (color online) Fourier transformed amplitude of EXAFS spectra for the $Zn_{1-x}Fe_xO$ (x=0.0, 0.01, 0.03, and 0.05) (a) Observed EXAFS spectra at Zn $K$-edge with Zn-foil for comparison. Inset: shows the points A and B in enhanced view. (b) Observed EXAFS spectra at Fe-$K$ edge with reference spectra FeO, $Fe_2O_3$ and $Fe_3O_4$ for comparison.

Figure 5: (color online) Fourier transformed EXAFS spectra for Fe doped ZnO at Zn $K$-edge data (dots) and simulated (lines). (a) shows real part of EXAFS theoretical spectra (lines) with experimental data (dots). (b) shows magnitude of the Fourier Transformed data (dots) and model fit (lines). The fit range is 1 to 6 Å.

# Table Caption

Table 1: The Linear Combination Fitting (LCF) results for Fe doped ZnO (x=0.01, 0.03 and 0.05) at Fe K-edge. R-factor and Chi-square values gives the fitting accuracy in assigned near-edge range (-10 - 50 eV).

Table 2: EXAFS fitting results for Zn K-edge spectra for the pure ZnO with Fe doped ZnO samples (x=0.0, 0.01, 0.03, and 0.05). While k-range=2.0-13.0 Å, R-range=1.0-6.0 Å, R-space fitting, dk=2, k-window = hanning, N in brackets is defined as co-ordination number (CN) or degeneracy. Scattering path is shown by # in the bracket of $1^{st}$ column.

Table 3: The values of EXAFS fitting parameters $E_0$, $\Delta R$, and uncertainties in k and R for Fe doped ZnO systems (x=0.0, 0.01, 0.03, and 0.05).

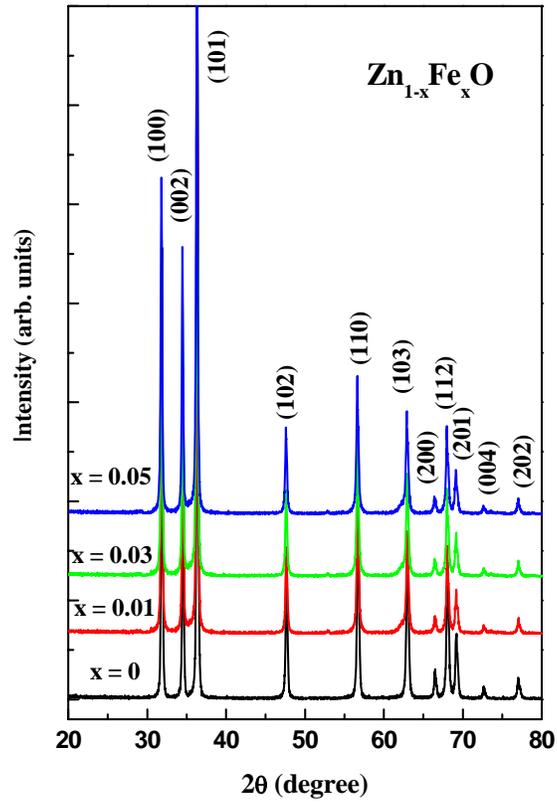

**Figure 1**

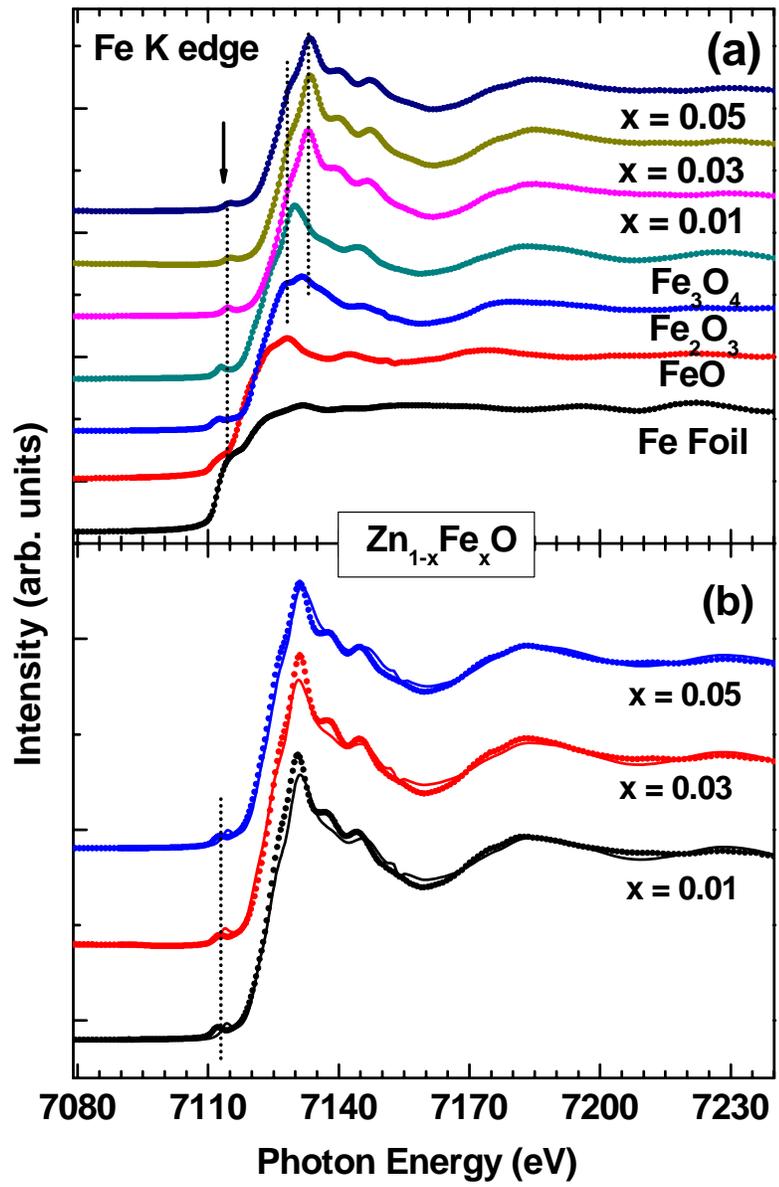

**Figure 2**

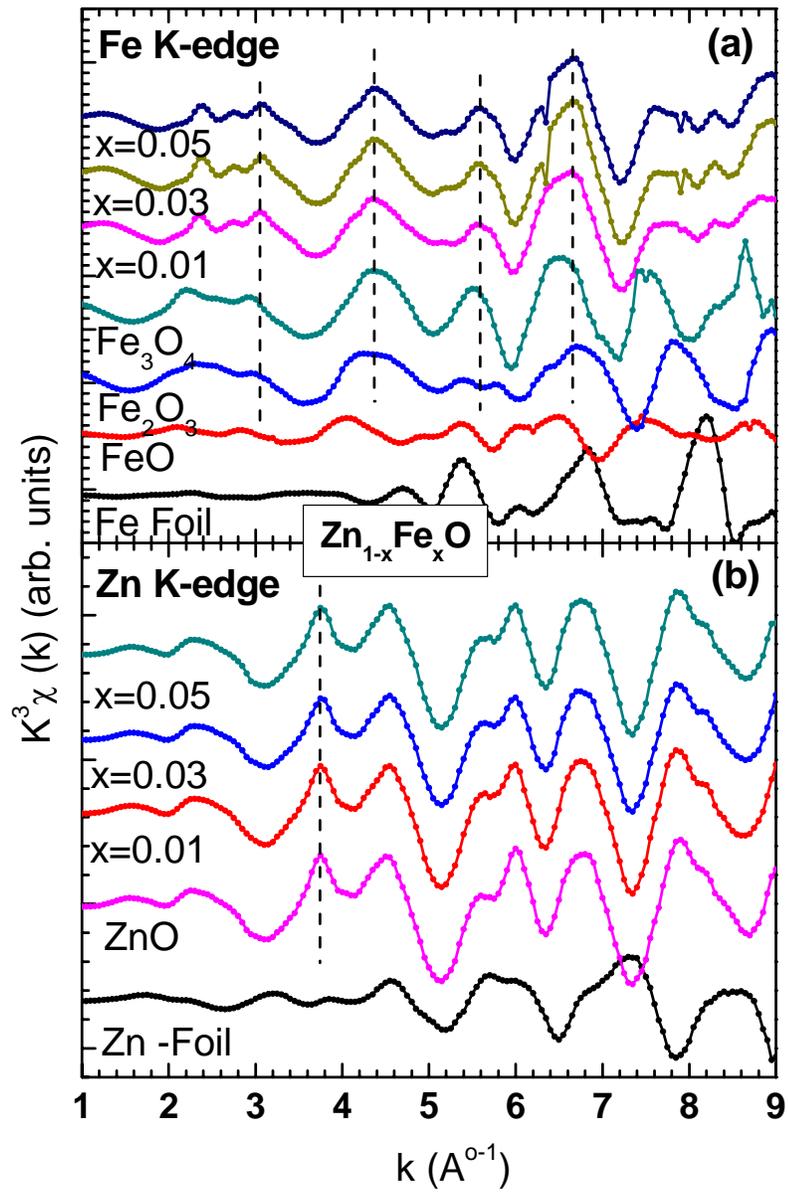

**Figure 3**

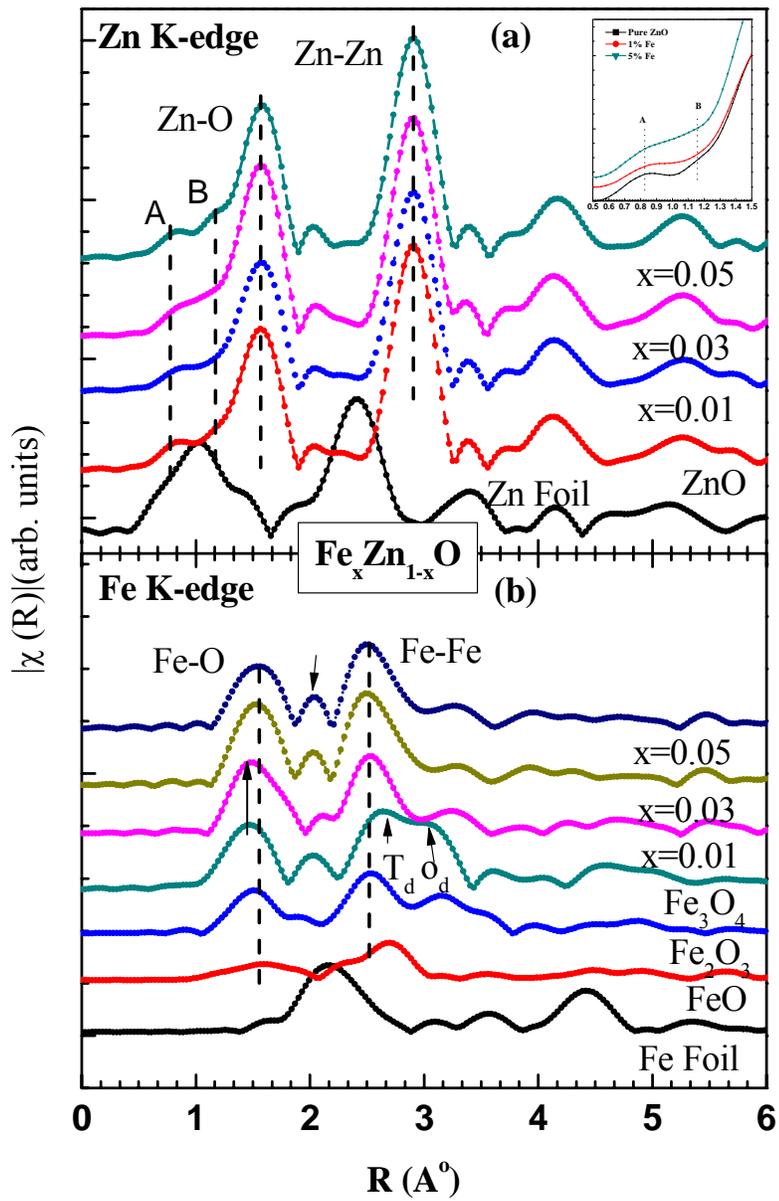

Figure 4

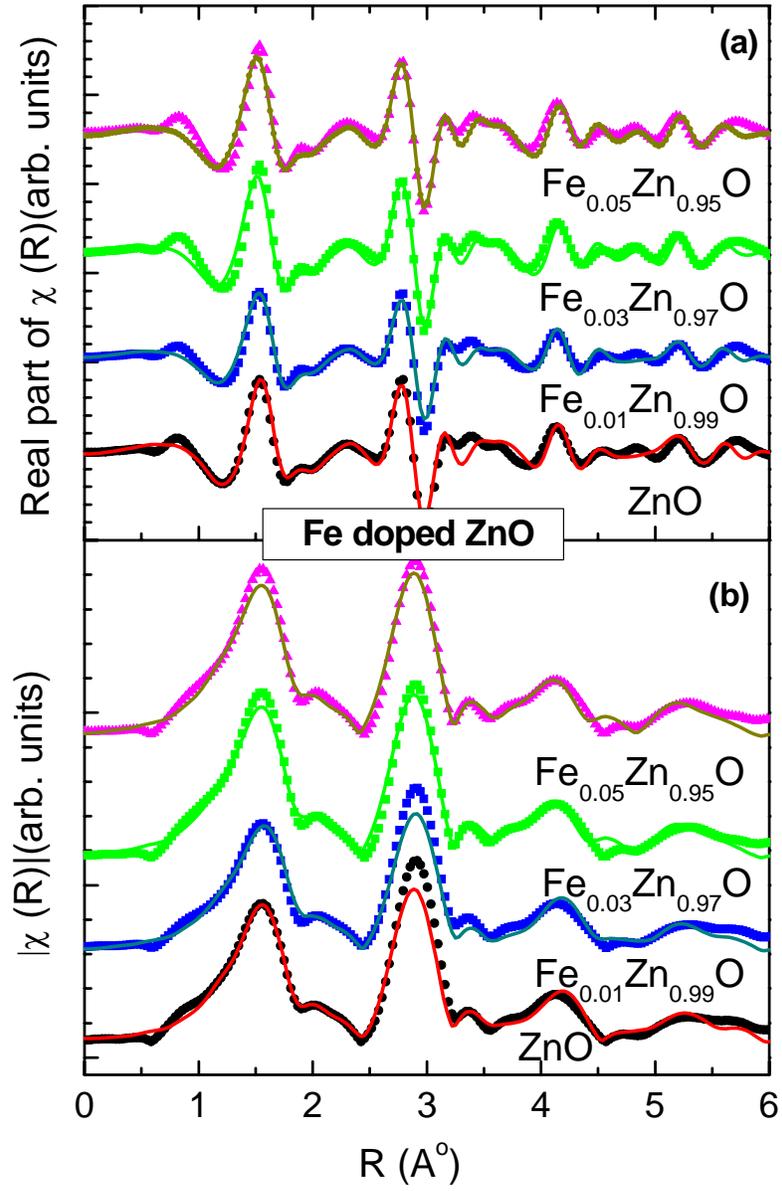

**Figure 5**

**Table 1**

| Sample | $Fe^{2+}$ | $Fe^{3+}$ | R-factor | Chi-square |
|---|---|---|---|---|
| $Fe_{0.01}Zn_{0.99}O$ | 0.3333 | 0.6667 | $6.49\times10^{-4}$ | $6.12\times10^{-2}$ |
| $Fe_{0.03}Zn_{0.97}O$ | 0.3333 | 0.6667 | $5.27\times10^{-4}$ | $4.64\times10^{-2}$ |
| $Fe_{0.05}Zn_{0.95}O$ | 0.2777 | 0.7223 | $4.03\times10^{-4}$ | $3.62\times10^{-2}$ |

**Table 2**

| Path | $R_{eff}$ (N)(Å) | $\sigma^2$ for x=0.0 | $\sigma^2$ for x=0.01 | $\sigma^2$ for x=0.03 | $\sigma^2$ for x=0.05 | $R_{eff}$ (N) (Å) |
|---|---|---|---|---|---|---|
| X-O$_1$ (#1) | 1.95(3) | 4.98 | 2.74 | 3.05 | 3.05 | 1.95 (3) for Zn-O$_1$ |
| X-O$_1$ (#2) | 2.03(1) | 4.98 | 2.74 | 3.05 | 3.05 | 1.99 (2) for X-O$_1$ |
| X-Zn$_2$ (#4) | 3.20(6) | 9.97 | 5.47 | 6.10 | 6.10 | 3.20 (6) for Zn-Zn$_1$ |
| X-Zn$_3$ (#5) | 3.25(3) | 9.97 | ---- | | ----- | 3.23 (4.5) for X–Zn$_1$ |
| X-O$_4$ (#11) | 3.79(3) | ----- | 5.47 | 6.10 | 6.10 | 3.83 (6) for Zn-O$_2$ |
| X-O$_5$ (#12) | 3.83(6) | 9.97 | 5.47 | 6.10 | 6.10 | 4.16 (5) for X-O$_2$ |
| X-O$_6$ (#25) | 4.53(6) | ------ | 5.47 | 6.10 | 6.10 | |
| X-Zn$_8$ (#28) | 4.56(6) | 9.97 | 5.47 | 6.10 | 6.10 | |
| X-O$_9$-Zn (#37) | 4.58(12) | ----- | 5.47 | 6.10 | 6.10 | |
| X-Zn$_{10}$-Zn (#38) | 4.83(24) | ----- | 5.47 | 6.10 | 6.10 | 4.56 (6) for Zn-Zn$_2$ |
| X-Zn$_{11}$-O (#39) | 4.84(12) | ----- | 5.47 | 6.10 | 6.10 | 4.75 (15) for X-Zn$_2$ |
| X-O$_{12}$-O (#40) | 4.84(12) | ----- | 5.47 | 6.10 | 6.10 | |
| X-O$_{13}$-Zn (#41) | 4.84(12) | ---- | 5.47 | 6.10 | 6.10 | |
| X-O$_{14}$ (#42) | 4.99(6) | 9.97 | 5.47 | 6.10 | 6.10 | 4.99 (6) for Zn-O$_3$ |
| X-O$_{15}$ (#43) | 5.00(3) | | 5.47 | 6.10 | 6.10 | 4.99 (4.5) for X-O$_3$ |

**Table 3**

| Fit/Parameters | ZnO | $Fe_{0.01}Zn_{0.99}O$ | $Fe_{0.03}Zn_{0.97}O$ | $Fe_{0.05}Zn_{0.95}O$ |
|---|---|---|---|---|
| $E_0$ (eV) | 5.68 | 6.63 | 5.80 | 5.83 |
| R-factor | 0.03032 | 0.05421 | 0.05200 | 0.05211 |
| ΔR | 0.01332 | 0.04118 | 0.03244 | 0.03555 |
| k-uncertainty (%) | 5.28 | 4.23 | 2.80 | 2.89 |
| R-uncertainty (%) | 9.35 | 9.11 | 8.80 | 8.85 |